\documentclass[letterpaper]{article} % DO NOT CHANGE THIS
\usepackage{aaai2026}  % DO NOT CHANGE THIS
\usepackage{times}  % DO NOT CHANGE THIS
\usepackage{helvet}  % DO NOT CHANGE THIS
\usepackage{courier}  % DO NOT CHANGE THIS
\usepackage[hyphens]{url}  % DO NOT CHANGE THIS
\usepackage{graphicx} % DO NOT CHANGE THIS
\usepackage{natbib}  % DO NOT CHANGE THIS AND DO NOT ADD ANY OPTIONS TO IT
\usepackage{caption} % DO NOT CHANGE THIS AND DO NOT ADD ANY OPTIONS TO IT
\usepackage{algorithm}
\usepackage{algorithmic}

\usepackage{amsmath}     % 基本数学环境和命令
\usepackage{amsfonts}    % 数学字体，包括 \mathbb{}
\usepackage{amssymb}     % 额外的数学符号
\usepackage{mathtools}   % amsmath的扩展

\usepackage{booktabs}       % professional-quality tables
\usepackage{colortbl}  %彩色表格需要加载的宏包
\usepackage{xcolor}         % colors
\usepackage{graphicx}  
\usepackage{arydshln} 

\usepackage{newfloat}
\usepackage{listings}
\DeclareCaptionStyle{ruled}{labelfont=normalfont,labelsep=colon,strut=off} % DO NOT CHANGE THIS
\floatstyle{ruled}
\newfloat{listing}{tb}{lst}{}
\floatname{listing}{Listing}
\title{Multi-dimensional Neural Decoding with Orthogonal Representations for Brain-Computer Interfaces}

\author{
Kaixi Tian\textsuperscript{1,2,3},
Shengjia Zhao\textsuperscript{4},
Yuhan Zhang\textsuperscript{1,2,3},
Shan Yu\textsuperscript{1,2,3}\thanks{Corresponding author}
}

\affiliations {
\textsuperscript{1}Laboratory of Brain Atlas and Brain-inspired Intelligence, Institute of Automation, Chinese Academy of Sciences\\
\textsuperscript{2}School of Future Technology, University of Chinese Academy of Sciences\\
\textsuperscript{3}State Key Laboratory of Brain Cognition and Brain-inspired Intelligence, CAS Center for Excellence in Brain Science and Intelligence Technology\\
\textsuperscript{4}Department of Physics, University of Oxford\\
tiankaixi2020@ia.ac.cn, shan.yu@nlpr.ia.ac.cn
}

\begin{document}

\maketitle

\begin{abstract}
Current brain-computer interfaces primarily decode single motor variables, limiting their ability to support natural, high-bandwidth neural control that requires simultaneous extraction of multiple correlated motor dimensions. We introduce Multi-dimensional Neural Decoding (MND), a task formulation that simultaneously extracts multiple motor variables (direction, position, velocity, acceleration) from single neural population recordings. MND faces two key challenges: cross-task interference when decoding correlated motor dimensions from shared cortical representations, and generalization issues across sessions, subjects, and paradigms. To address these challenges, we propose \textbf{OrthoSchema}, a multi-task framework inspired by cortical orthogonal subspace organization and cognitive schema reuse. OrthoSchema enforces representation orthogonality to eliminate cross-task interference and employs selective feature reuse transfer for few-shot cross-session, subject and paradigm adaptation. Experiments on macaque motor cortex datasets demonstrate that OrthoSchema significantly improves decoding accuracy in cross-session, cross-subject and challenging cross-paradigm  generalization tasks, with larger performance improvements when fine-tuning samples are limited. Ablation studies confirm the synergistic effects of all components are crucial, with OrthoSchema effectively modeling cross-task features and capturing session relationships for robust transfer. Our results provide new insights into scalable and robust neural decoding for real-world BCI applications.
\end{abstract}


\section{Introduction}
Brain-computer interfaces (BCIs) represent a transformative technology for restoring motor function in patients with neural injuries, providing opportunities for decoding neural signals and translating them into actionable commands. However, traditional BCI systems have focused on single-output decoding tasks, extracting individual motor variables from motor cortex \citep{Hochberg2012,Flesher2021}. Natural motor control involves coordinated multi-dimensional information processing, where the brain simultaneously encodes multi-dimensional representations within the same cortical populations \citep{Cunningham2014,Li2014}. Consequently, current BCI systems still struggle to support natural, fluent interactions that require information-rich, high-dimensional neural representations\citep{Collinger2013,Wolpaw2013}, resulting in slow decoding responses and poor generalization in real-world scenarios \citep{Shih2012}. We propose Multi-dimensional Neural Decoding (MND) as a novel task formulation that simultaneously extracts multiple correlated motor variables from shared neural population activity. Unlike traditional single-task approaches, MND formulation encounters the complex interactions that arise when decoding correlated motor dimensions from shared cortical representations. This task formulation not only captures the biological reality of cortical motor encoding but also meets the growing demand for high-bandwidth BCI control \citep{Gallego2022}.

\begin{figure}[t]
\centering
\includegraphics[width=\columnwidth]{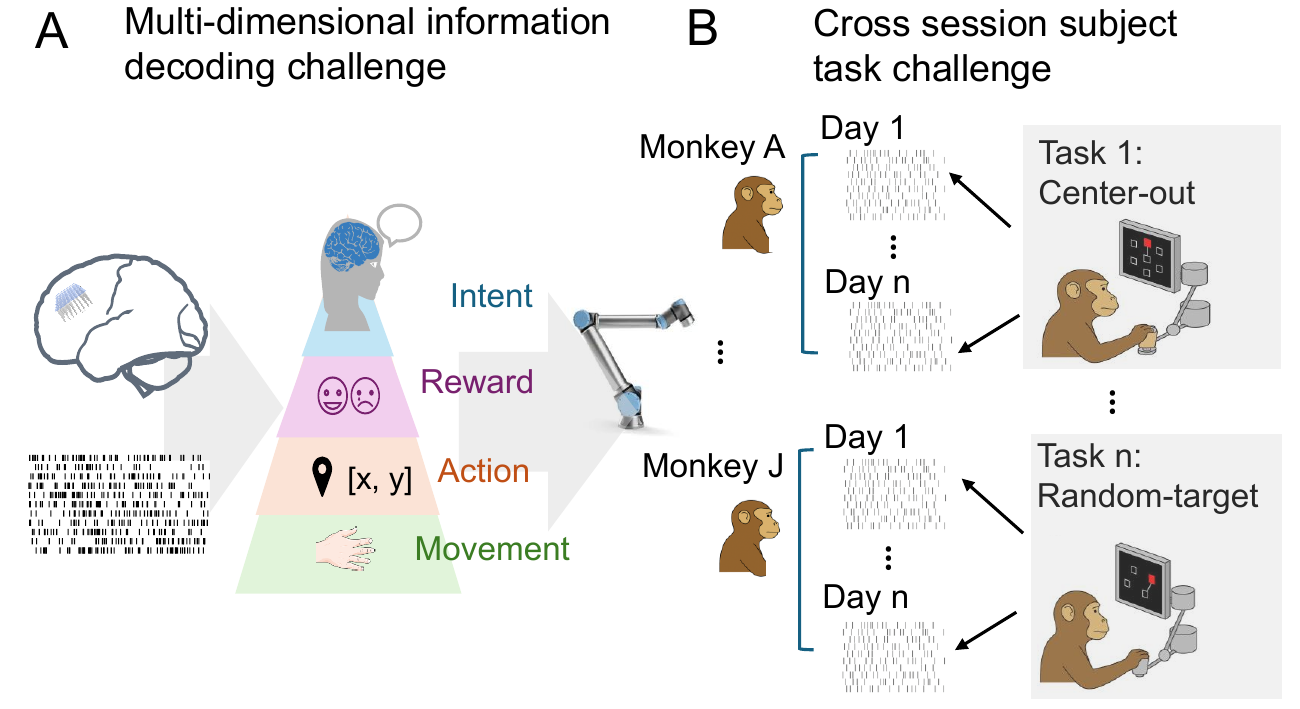} % Reduce the figure size so that it is slightly narrower than the column. Don't use precise values for figure width.This setup will avoid overfull boxes.
\caption{Key challenges in multi-dimensional neural decoding. (A) Growing demand for parallel decoding of diverse neural signals. (B) Limited generalization under distribution shifts across sessions, subjects, and paradigms.}
\label{fig1}
\end{figure} 

MND faces two core challenges, as illustrated in Figure \ref{fig1}. First, it remains unclear whether sufficient multi-dimensional information across different hierarchical levels can be extracted from limited implanted brain regions. When extending to multi-dimensional scenarios, severe cross-task interference exists among different dimensions, where extraction of one type of information adversely affects the decoding accuracy of other dimensions \citep{10.7554/eLife.87881}. Second, since neural signals naturally drift over time \citep{Karpowicz2025,Wimalasena2020}, existing methods \citep{Lim2025} exhibit poor adaptability across experimental sessions, subjects, and paradigm variations, often requiring extensive retraining when facing new conditions.

Neural coding research provides important insights for addressing these challenges. Studies in rodents \citep{Zhou2021} and non-human primates \citep{Bernardi2020} demonstrate that the brain can form schema-like representational patterns that remain consistent across sessions and subjects. Meanwhile, non-human primate research reveals that different functional representations within the same cortical area exhibit orthogonal properties \citep{Tian2024, Flesch2022}, and this orthogonal subspace organization facilitates interference-free representation of multi-dimensional information. These findings inspire two insights for algorithm design: orthogonal subspace organization may offer a promising approach to mitigate feature coupling problems in multi-dimensional decoding, while schema-like stable representation could provide a theoretical foundation for more efficient transfer learning strategies.

\begin{figure}[t]
\centering
\includegraphics[width=\columnwidth]{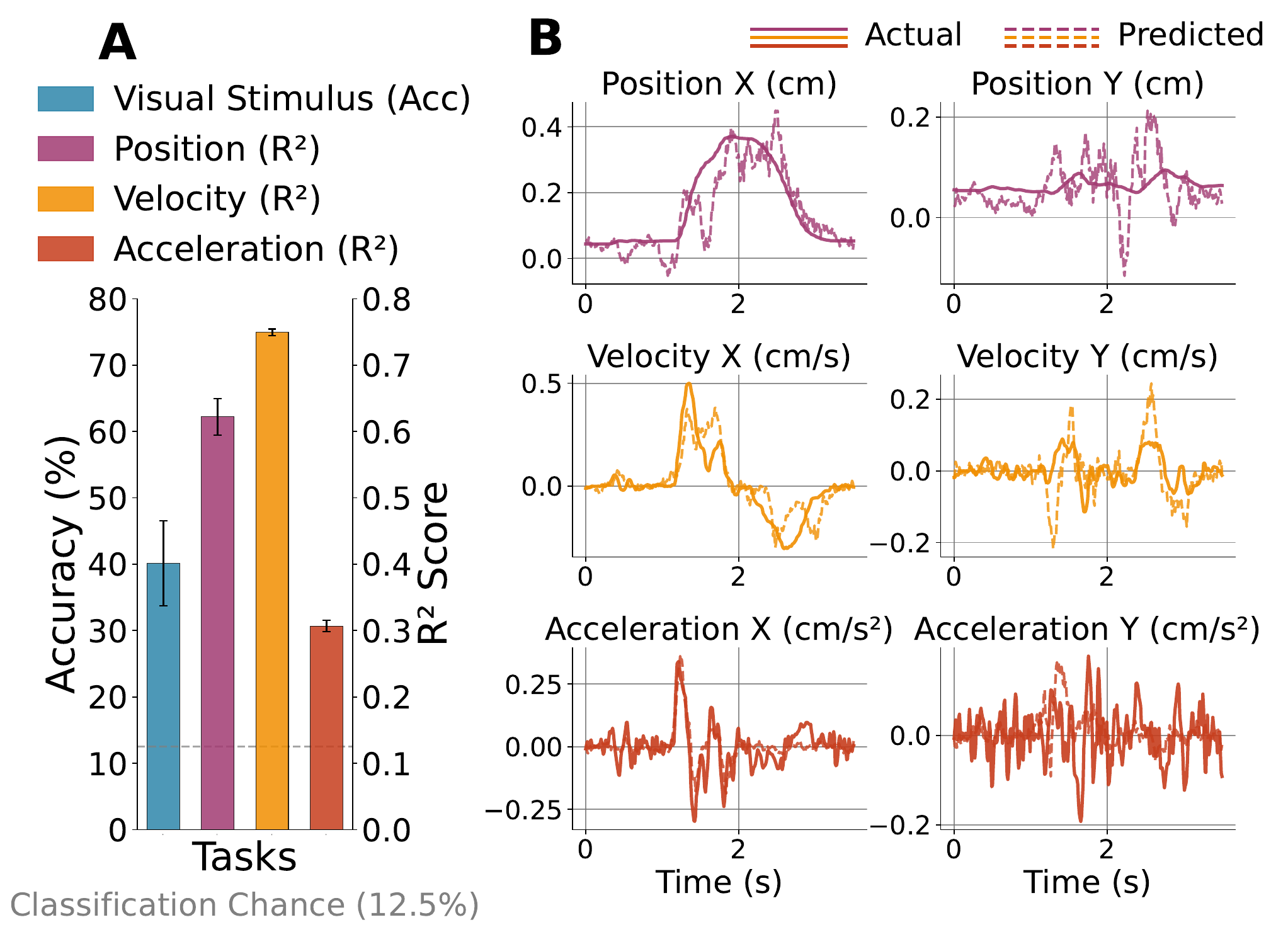} % Reduce the figure size so that it is slightly narrower than the column. Don't use precise values for figure width.This setup will avoid overfull boxes.
\caption{Feasibility of decoding distinct motor features from cortical activity. (A) Independent decoding of multiple motor features (direction, position, velocity, and acceleration) from motor cortex recordings within the same session. (B) Sample predictions for hand motor variables.}
\label{fig2}
\end{figure} 

Drawing on these insights, we propose \textbf{OrthoSchema}—a structured multi-task neural decoding framework that aims to extract independent latent representations from neural activity and achieve improved generalization across subjects, sessions, and paradigms. The framework comprises three main components: (1) orthogonality constraints in the latent space to enhance separation among task-relevant features and reduce interference across decoding objectives; (2) an explicit session modeling mechanism that captures variations across sessions and subjects, with these representations excluded during transfer to remove drift-related information; (3) a parameter transfer strategy guided by feature relevance, enabling efficient few-shot adaptation to new sessions, subjects, or paradigms.

To validate the effectiveness of our approach, we employed macaque motor cortex neural datasets from two different paradigms: center-out reaching and random target tasks \citep{Perich2025}. We first experimentally demonstrated the practical feasibility of independently decoding multiple motor features from motor cortex (M1), as shown in Figure \ref{fig2}, confirming the potential for multi-task neural decoding. Subsequently, we evaluated the few-shot capabilities of the OrthoSchema architecture across different backbone networks in cross-subject, cross-session, and cross-paradigm scenarios. Experimental results demonstrate that our method achieves significant performance improvements across all testing scenarios, with particularly outstanding performance in the extremely few-shot regime (K=2). Ablation studies and visualization analyses suggest that orthogonal constraints promote the formation of task-specific feature patterns, while the schema-based transfer mechanism successfully decouples task-relevant representations from session-specific variations, providing mechanistic insights into the model's superior generalization capability. The main contributions of this work include:

\begin{itemize}

\item We propose MND as a BCI task formulation that meets the growing demand in BCI systems for simultaneously decoding multiple correlated motor variables from the same neural population.

\item We develop the \textbf{OrthoSchema} framework, combining orthogonal constraints and schema-based transfer mechanisms to effectively address cross-task interference and cross-session/subject/paradigm generalization challenges.

\item We validate the effectiveness of our framework on macaque motor cortex datasets, demonstrating superior few-shot transfer performance across cross-session, subject, and paradigm scenarios. 

\end{itemize}

\begin{figure*}[t]
\centering
\includegraphics[width=0.95\textwidth]{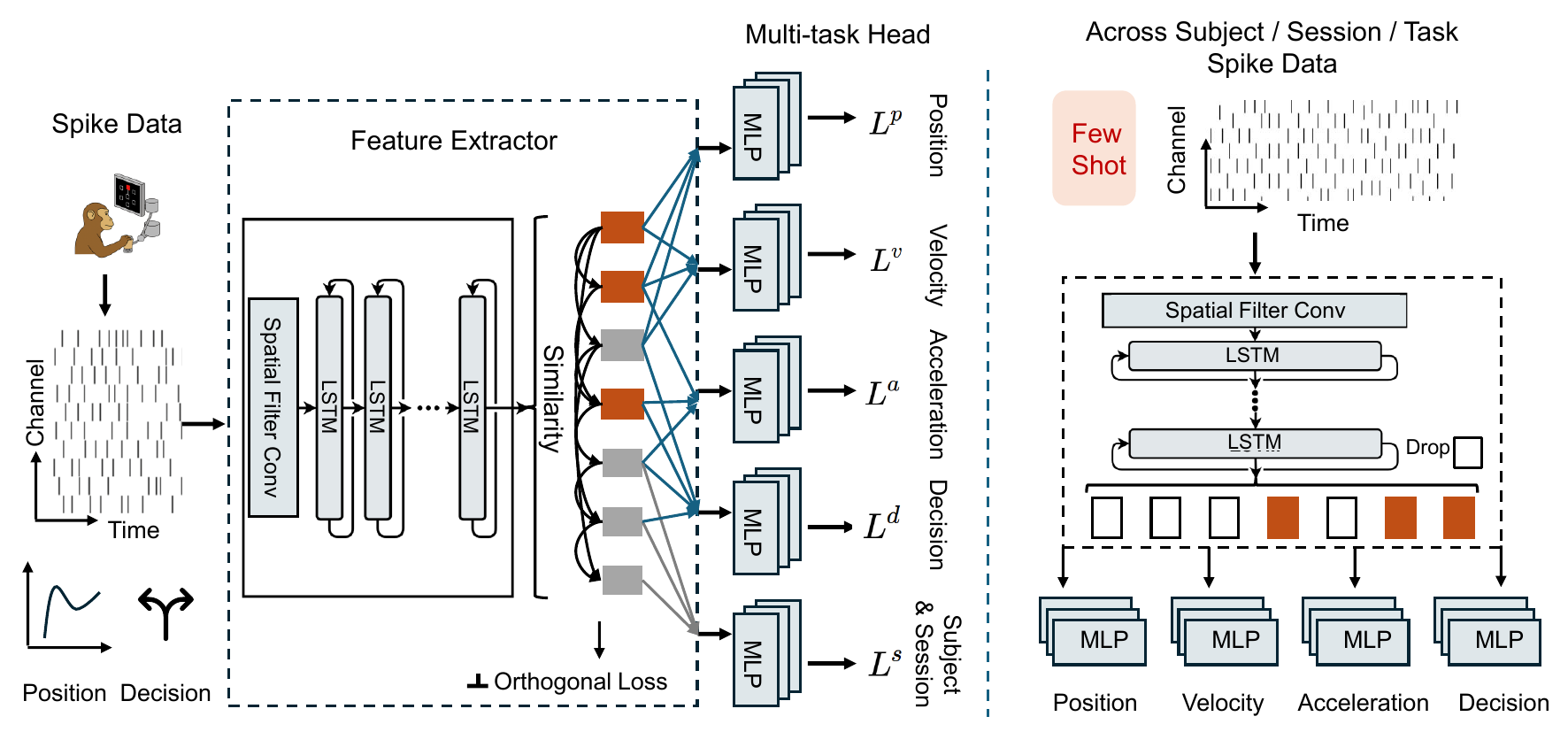} % Reduce the figure size so that it is slightly narrower than the text width. This setup will avoid overfull boxes.
\caption{Overview of the OrthoSchema framework. Spike data are processed by a global convolution layer followed by LSTM blocks (replaceable) to obtain latent representations. Orthogonality constraints are applied in the latent space. The model includes multiple decoding heads for direction classification and the regression of hand position, velocity, and acceleration. A session/subject classification head is employed during training to model distributional shifts. During inference, the session classification components are removed, and parameters are selectively reused for few-shot fine-tuning.}
\label{fig3}
\end{figure*}

\section{Related Work}
\subsection{Neural Decoding} 
Neural decoding refers to the computational process of extracting behavioral intentions and cognitive states from recorded neural signals \citep{Musallam2004}. Traditional BCI decoding methods primarily rely on linear models, such as Kalman filters \citep{Wu2006} and population vector algorithms \citep{Georgopoulos1986}, for motor intention prediction. Deep learning techniques have significantly improved decoding capabilities \citep{Glaser2020}. LSTM networks have been applied to motion trajectory decoding \citep{Liu2022}, while convolutional neural networks have enhanced spatial feature extraction \citep{Xie2018}. Recent approaches focus on modeling population neural dynamics. LFADS \citep{Pandarinath2018} uses variational autoencoders for latent neural state inference. NOMAD \citep{Karpowicz2025} improves motor decoding through latent variable alignment. However, these methods typically address single decoding objectives and face significant challenges when simultaneously extracting multiple behavioral variables from the same neural population.

\subsection{Multi-dimensional Information Decoding} With the demand for richer interaction capabilities in BCI applications, multi-dimensional information decoding has emerged as a central challenge in current neural decoding methods \citep{Wolpaw2013}. MND has been applied to EEG-based decoding, including Bayesian multi-task models \citep{Alamgir2010,Oikonomou2022} and deep neural networks with multi-head outputs \citep{Bertalan2021}, demonstrating cross-dimensional interference challenges.

Existing multi-task learning methods address task conflicts through  shared/private subspace decomposition \citep{dong2023gdod}, adaptive loss weighting via gradient magnitudes \citep{Chen2018GradNorm}, or uncertainty-based task weighting \citep{Kendall_2018_CVPR}. However, neuroscience research suggests that neural population can form mutually orthogonal encodings in output subspaces \citep{Kaufman2014, Tang2020Minimally, Elsayed2016}, encouraging features to be approximately orthogonal without interfering with the shared encoder's general representation capability.

\subsection{Cross-session/subject/paradigm Adaptation} Neural decoding stability across sessions, subjects, and paradigms poses major challenges for practical BCI deployment, as neural activity changes significantly due to electrode displacement and tissue reactions \citep{Perge2013,Chestek2011,Willett2021}. \citet{farshchian2018adversarial} proposed adversarial learning-based domain adaptation methods that reduce cross-session variability by learning domain-invariant representations. \citet{Degenhart2020} developed alignment strategies based on shared neural manifolds, achieving cross-session generalization. Recent foundation models such as Neural Data Transformer \citep{Ye2023} and POYO \citep{Ye2024a} achieve cross-subjects generalization through large-scale pretraining. However, these methods require extensive neural data for pretraining while ignoring the differential stability of schema-like neural representations \citep{Goudar2023Schema} across sessions, subjects, and paradigms revealed in neuroscience research.

Unlike existing single-output decoders, OrthoSchema is specifically designed for MND tasks, integrating orthogonality constraints in latent space with selective schema reuse to enhance multi-dimensional decoding stability across sessions, subjects, and paradigms. 

\section{Method}
MND is defined as the simultaneous extraction of multiple correlated motor variables $\{y_1, y_2, ..., y_n\}$ from shared neural population recordings $X$, where variables exhibit hierarchical dependencies (e.g., position, velocity, acceleration). We propose OrthoSchema to address MND's core challenges: cross-task interference from correlated dimensions and poor generalization across sessions, subjects, and paradigms through orthogonal feature disentanglement and schema-based transfer learning.

Figure~\ref{fig3} illustrates OrthoSchema neural decoding framework comprising four modules: (a) orthogonally-constrained feature extraction; (b) multi-decoding heads; (c) transfer with selective feature reuse. The pipeline processes neural spike data through spatial filtering and hierarchical networks, applies orthogonal constraints for feature diversity, employs multi-task learning for BCI decoding and session classification, and achieves few-shot generalization through selective parameter transfer.

\subsection{Orthogonally-Constrained Feature Extractor}
The orthogonally-constrained feature extractor extracts decoupled features from neural spike data to reduce cross-dimensional interference. This module employs a model-agnostic design with a scalable shared encoder that can be adapted to different backbone architectures. Neural spike data $X \in \mathbb{R}^{B \times 1 \times C \times T}$ (where $B$ is batch size, $C$ is channels, $T$ is time steps) is first processed through a spatial filter to extract coordinated neural activity patterns:
\begin{equation}
h_0 = \text{ELU}(\text{BatchNorm2d}(\text{Conv2d}(X, \theta_0)))
\end{equation}
where $\text{Conv2d}(\cdot; \theta_0)$ denotes 2D convolution with learnable parameters $\theta_0$, $\text{BatchNorm2d}(\cdot)$ represents batch normalization for stable training, and $\text{ELU}(\cdot)$ is the Exponential Linear Unit activation function. After dimension adjustment, the data is processed through bidirectional LSTM layers:
\begin{equation}
h_{k+1}, (c_{k+1}, m_{k+1}) = \text{LSTM}_k(h_k, (c_k, m_k))
\end{equation}
\begin{equation}
h_{k+1} = \text{ELU}(\text{BatchNorm1d}(h_{k+1}^T))^T
\end{equation}
where $k = 1, 2, \ldots, 6$, $h_k$ represents the hidden state, $(c_k, m_k)$ denote the cell state and memory state of the LSTM at layer $k$. 

\paragraph{Orthogonal Feature Constraint} We impose orthogonal constraints on the final RNN layer features to enhance feature diversity and promote independent representations across-dimensions.
The orthogonality loss is computed as:
\begin{equation}
L_{\text{orthogonal}} = ||\text{mean}(x_{norm} \cdot x_{norm}^T) - I||_F^2
\end{equation}
where $x_{norm}$ represents the normalized features, $I$ is the identity matrix.

\subsection{Multi-Task Head} 
 All task heads connect to the final RNN features $h_L$ processed with orthogonal constraints. The architecture includes two types of task heads: (1) BCI decoding heads for motor-related variables and (2) a session classification head to model subject/session-specific features and promote cross-session generalization.

Each BCI decoding head is implemented as a multi-layer perceptron (MLP), targeting a specific aspect of motor behavior. For direction classification, we use temporally averaged features:
\begin{equation}
\bar{x} = \frac{1}{T}\sum_{t=1}^{T} x^{(t)} \in \mathbb{R}^{B \times 128}
\end{equation}
\begin{equation}
f_{\text{direction}} = \text{MLP}_{\text{direction}}(\bar{x}) \in \mathbb{R}^{B \times 8}
\end{equation}
where $x^{(t)}$ represents the feature at the $t$-th time step, and $\bar{x}$ denotes the temporally averaged features. For the three motor regression tasks:
\begin{align}
f_{\text{position}} &= \text{MLP}_{\text{position}}(x), \\ f_{\text{velocity}} &= \text{MLP}_{\text{velocity}}(x) \\
f_{\text{acceleration}} &= \text{MLP}_{\text{acceleration}}(x)
\end{align}
where $x \in \mathbb{R}^{B \times T \times 128}$ represents the features for time-step-wise processing.

To identify and model session/subject drift information, we design a classification task head to recognize neural signal characteristics across different subjects and sessions:
\begin{equation}
f_{\text{session}} = \text{MLP}_{\text{session}}(x_{avg}) \in \mathbb{R}^{B \times N_{sessions}}
\end{equation}
where $f_{\text{session}}$ represents the output probabilities for session classification, $x_{avg}$ denotes the temporally averaged features, and $N_{sessions}$ is the total number of sessions in the training data. This task head helps capture session/subject-related features and enables identification of stable representations by filtering out session/subject-specific components during few-shot transfer.

\paragraph{Loss Function} The total loss function integrates losses from all tasks along with the orthogonal constraint loss:
\begin{align}
L_{\text{total}} = &\alpha_1 L_{\text{dir}} + \alpha_2 (L_{\text{pos}} + L_{\text{vel}} + L_{\text{acc}}) \nonumber \\
&+ \alpha_3 L_{\text{orthogonal}} + \alpha_4 L_{\text{session}}
\end{align}
where classification tasks employ cross-entropy loss and regression tasks use mean squared error loss. $\alpha_1, \alpha_2, \alpha_3, \alpha_4$ are weighting factors that balance different objectives.

\subsection{Transfer with Selective Feature Reuse}
Inspired by the brain's ability to reuse stable cognitive schemas for cross-session, subject and paradigm generalization \citep{samborska2022complementary}, we develop a selective parameter transfer strategy for efficient cross-session/subject/paradigm adaptation. Our approach focuses on the final layer of RNN features and computes task-relevant and session/subject-specific relevance separately:

We first compute the average weight importance across all task heads to identify shared representations for multiple motor decoding tasks:
\begin{equation}
I_{\text{task}}(\theta_i) = \frac{1}{|H|}\sum_{h \in H} |\text{weight}_{h,i}|
\end{equation}
where $H$ is the set of BCI task heads and $\text{weight}_{h,i}$ is the weight connecting the $i$-th feature to task head $h$. Simultaneously, we compute the parameter importance for the session classification task to capture session/subject-specific drift features:
\begin{equation}
I_{\text{session}}(\theta_i) = |\text{weight}_{\text{session},i}|
\end{equation}
where $\text{weight}_{\text{session},i}$ is the weight connecting the $i$-th feature to the session classification head.

We extract task-relevant yet stable representations through importance difference:
\begin{equation}
I_{\text{transfer}}(\theta_i) = I_{\text{task}}(\theta_i) - \beta \cdot I_{\text{session}}(\theta_i)
\end{equation}
\begin{equation}
\Theta_{\text{transfer}} = \{\theta_i | I_{\text{transfer}}(\theta_i) \geq \text{percentile}(I_{\text{transfer}}, 100-\gamma)\}
\end{equation}
where $\beta$ controls the degree of down-weighting session-specific features, and $\gamma$ is the percentage of top-ranked features to select. This strategy ensures that selected features are important for BCI tasks but contribute little to session classification. During k-shot fine-tuning, we only update the selected parameters in $\Theta_{\text{transfer}}$, ensuring that only task-relevant stable features receive gradient updates for effective few-shot generalization.

\begin{table*}[!ht]
\centering
\scriptsize
\renewcommand{\arraystretch}{1.2}
\setlength{\tabcolsep}{2.5pt}
\begin{tabular}{llcccccccc}
\toprule
\textbf{Dataset} & \textbf{Methods} & \multicolumn{8}{c}{\textbf{Across Session}} \\
\cmidrule(lr){3-10}
& & \multicolumn{2}{c}{\textbf{Direction (\%)}} 
  & \multicolumn{2}{c}{\textbf{Hand Position ($R^2$)}} 
  & \multicolumn{2}{c}{\textbf{Hand Velocity ($R^2$)}} 
  & \multicolumn{2}{c}{\textbf{Hand Acceleration ($R^2$)}} \\
\cmidrule(lr){3-4} \cmidrule(lr){5-6} \cmidrule(lr){7-8} \cmidrule(lr){9-10}
& & K=2 & K=5 & K=2 & K=5 & K=2 & K=5 & K=2 & K=5 \\
\midrule

& CNN   & $33.97 \pm 3.93$ & $73.55 \pm 6.11$ & $0.533 \pm 0.080$ & $0.724 \pm 0.038$ & $0.650 \pm 0.022$ & $0.783 \pm 0.024$ & $0.299 \pm 0.147$ & $0.562 \pm 0.049$ \\
& CNN+OS & $\mathbf{37.99 \pm 1.35}$ & $\mathbf{79.42 \pm 2.24}$ & $\mathbf{0.644 \pm 0.033}$ & $\mathbf{0.769 \pm 0.017}$ & $\mathbf{0.720 \pm 0.017}$ & $\mathbf{0.801 \pm 0.019}$ & $\mathbf{0.440 \pm 0.020}$ & $0.550 \pm 0.014$ \\
\cdashline{2-10}[1pt/2pt]
& RNN   & $72.60 \pm 4.64$ & $88.32 \pm 0.53$ & $0.886 \pm 0.015$ & $0.950 \pm 0.002$ & $0.838 \pm 0.013$ & $0.916 \pm 0.003$ & $0.576 \pm 0.015$ & $0.715 \pm 0.017$ \\
{\textbf{CO}} & RNN+OS & $\mathbf{80.23 \pm 0.51}$ & $\mathbf{89.68 \pm 0.41}$ & $\mathbf{0.903 \pm 0.005}$ & $\mathbf{0.952 \pm 0.004}$ & $\mathbf{0.862 \pm 0.014}$ & $\mathbf{0.922 \pm 0.002}$ & $\mathbf{0.611 \pm 0.004}$ & $\mathbf{0.735 \pm 0.010}$ \\
\cdashline{2-10}[1pt/2pt]
& MLP   & $/$ & $94.02 \pm 2.15$ & $/$ & $0.860 \pm 0.009$ & $/$ & $0.529 \pm 0.055$ & $/$ & $/$ \\
& MLP+OS & $/$ & $\mathbf{94.39 \pm 0.86}$ & $/$ & $\mathbf{0.863 \pm 0.006}$ & $/$ & $\mathbf{0.554 \pm 0.006}$ & $/$ & $/$ \\

\midrule

& CNN   & $-$ & $-$ & $0.566 \pm 0.034$ & $0.708 \pm 0.022$ & $0.698 \pm 0.017$ & $0.774 \pm 0.022$ & $-$ & $-$ \\
& CNN +OS & $-$ & $-$ & $\mathbf{0.601 \pm 0.022}$ & $\mathbf{0.727 \pm 0.032}$ & $\mathbf{0.719 \pm 0.007}$ & $\mathbf{0.798 \pm 0.012}$ & $-$ & $-$ \\
\cdashline{2-10}[1pt/2pt]
{\textbf{RT}}& RNN   & $-$ & $-$ & $0.635 \pm 0.017$ & $0.712 \pm 0.057$ & $0.694 \pm 0.008$ & $0.782 \pm 0.008$ & $-$ & $-$ \\
& RNN +OS & $-$ & $-$ & $\mathbf{0.681 \pm 0.019}$ & $\mathbf{0.768 \pm 0.050}$ & $\mathbf{0.717 \pm 0.014}$ & $\mathbf{0.794 \pm 0.020}$ & $-$ & $-$ \\
\bottomrule
\end{tabular}
\caption{
Cross-session performance under 2-shot and 5-shot settings (mean±std). Direction is reported as accuracy (\%), while position, velocity, and acceleration are reported as $R^2$ values. Bold values represent the best results within each K-shot group. "/" indicates results at chance level. "-" indicates the task is not applicable. +OS denotes enhancement with OrthoSchema.
}
\label{tab:across_session}
\end{table*}

\begin{table}[!ht]
\centering
\scriptsize
\renewcommand{\arraystretch}{1.2}
\setlength{\tabcolsep}{4pt}
\begin{tabular}{llccc}
\toprule
\textbf{Methods} & \textbf{K-shot} & \multicolumn{3}{c}{\textbf{Across Subject}} \\
\cmidrule(lr){3-5}
& & \textbf{Direction (\%)} & \textbf{Position ($R^2$)} & \textbf{Velocity ($R^2$)} \\
\midrule
CNN & K=2 & $38.93 \pm 6.53$ & $0.504 \pm 0.049$ & $0.538 \pm 0.026$ \\
CNN+OS & K=2 & $\mathbf{44.76 \pm 2.84}$ & $\mathbf{0.546 \pm 0.039}$ & $\mathbf{0.543 \pm 0.012}$ \\
\cdashline{1-5}[1pt/2pt]
CNN & K=5 & $65.52 \pm 4.31$ & $0.620 \pm 0.044$ & $0.622 \pm 0.015$ \\
CNN+OS & K=5 & $\mathbf{66.50 \pm 5.41}$ & $\mathbf{0.663 \pm 0.040}$ & $\mathbf{0.644 \pm 0.018}$ \\
\cdashline{1-5}[1pt/2pt]
RNN & K=2 & $45.53 \pm 4.45$ & $0.685 \pm 0.039$ & $0.592 \pm 0.017$ \\
RNN+OS  & K=2 & $\mathbf{60.01 \pm 3.69}$ & $\mathbf{0.783 \pm 0.022}$ & $\mathbf{0.653 \pm 0.021}$ \\
\cdashline{1-5}[1pt/2pt]
RNN & K=5 & $75.65 \pm 4.43$ & $0.824 \pm 0.017$ & $0.697 \pm 0.022$ \\
RNN+OS & K=5 & $\mathbf{77.76 \pm 1.97}$ & $\mathbf{0.872 \pm 0.009}$ & $\mathbf{0.743 \pm 0.011}$ \\
\cdashline{1-5}[1pt/2pt]
MLP & K=5 & $90.25 \pm 1.46$ & $0.854 \pm 0.008$ & $0.589 \pm 0.020$ \\
MLP+OS & K=5 & $\mathbf{91.02 \pm 0.71}$ & $\mathbf{0.856 \pm 0.008}$ & $\mathbf{0.607 \pm 0.004}$ \\
\bottomrule
\end{tabular}
\caption{
Cross-subject performance on CO task under 2-shot and 5-shot settings (mean±std). Direction is reported as accuracy (\%), while position, velocity are reported as $R^2$ values. Bolded values are the best results within each K-shot group. +OS indicates enhanced with OrthoSchema.
}
\label{tab:across_subject}
\end{table}

\begin{table}[!ht]
\centering
\scriptsize
\renewcommand{\arraystretch}{1.2}
\setlength{\tabcolsep}{6pt}
\begin{tabular}{llcc}
\toprule
\textbf{Methods} & \textbf{K-shot} & \multicolumn{2}{c}{\textbf{Across Paradigm (CO$\rightarrow$RT)}} \\
\cmidrule(lr){3-4}
& & \textbf{Position ($R^2$)} & \textbf{Velocity ($R^2$)} \\
\midrule
CNN & K=2 & $0.233 \pm 0.061$ & $0.400 \pm 0.046$ \\
CNN+OS & K=2 & $\mathbf{0.240 \pm 0.060}$ & $\mathbf{0.427 \pm 0.026}$ \\
\cdashline{1-4}[1pt/2pt]
CNN & K=5 & $0.359 \pm 0.038$ & $0.531 \pm 0.031$ \\
CNN+OS & K=5 & $\mathbf{0.425 \pm 0.024}$ & $\mathbf{0.546 \pm 0.011}$ \\
\cdashline{1-4}[1pt/2pt]
RNN & K=2 & $0.579 \pm 0.040$ & $0.688 \pm 0.023$ \\
RNN+OS & K=2 & $\mathbf{0.622 \pm 0.013}$ & $\mathbf{0.697 \pm 0.000}$ \\
\cdashline{1-4}[1pt/2pt]
RNN & K=5 & $0.707 \pm 0.013$ & $0.762 \pm 0.006$ \\
RNN+OS & K=5 & $\mathbf{0.728 \pm 0.009}$ & $\mathbf{0.782 \pm 0.003}$ \\
\cdashline{1-4}[1pt/2pt]
MLP & K=5 & $0.618 \pm 0.018$ & $0.250 \pm 0.015$ \\
MLP+OS & K=5 & $\mathbf{0.624 \pm 0.002}$ & $\mathbf{0.257 \pm 0.010}$ \\
\bottomrule
\end{tabular}
\caption{
Cross-paradigm performance (CO→RT) under 2-shot and 5-shot settings (mean±std).  Position and Velocity are reported as $R^2$ values. Bolded values are the best results. +OS indicates enhanced with OrthoSchema.
}
\label{tab:across_task}
\end{table}

\section{Experiments}
\subsection{CO and RT Dataset}

We use motor cortex neural datasets from three rhesus macaques \cite{Perich2025}. The dataset contains two motor paradigms: center-out (CO) tasks require macaques to perform reaching movements from the center position to the targets based on one of directional cues displayed on the screen; random target (RT) tasks require macaques to reach targets that appear at random locations on the screen. Subject C provided 14 CO task sessions (recorded September-October 2016), Subject M provided 20 CO task sessions and 6 RT task sessions (recorded January 2014-June 2015), and Subject T provided 6 CO task sessions and 6 RT task sessions (recorded August-September 2013). These data provide a testing foundation for validating the model's cross-session, cross-subject, and cross-paradigm generalization capabilities in MND. CO tasks contain neural spike activity, hand position, velocity, acceleration, and directional choice labels; RT tasks contain neural spike activity, hand position, and velocity. For neural data processing, we extract $T = 3.5$ seconds of neural activity starting from target onset with sampling rate $f_s = 100$ Hz. Additional implementation details are provided in the appendix.

\begin{table}[!ht]
\centering
\scriptsize
\renewcommand{\arraystretch}{1.2}
\setlength{\tabcolsep}{1pt}
\resizebox{\columnwidth}{!}{%
\begin{tabular}{lcccc}
\toprule
\textbf{Methods} & \textbf{Direction (\%)} & \textbf{Position ($R^2$)} & \textbf{Velocity ($R^2$)} & \textbf{Acceleration ($R^2$)} \\
\midrule
\multicolumn{5}{c}{K=2} \\
\midrule
RRR & $37.60 \pm 4.20$ & $0.456 \pm 0.066$ & $0.102 \pm 0.037$ & $-0.260 \pm 0.062$ \\
LFADS & $64.88 \pm 4.35$ & $0.841 \pm 0.040$ & $0.818 \pm 0.019$ & $0.555 \pm 0.021$ \\
UW & $52.03 \pm 1.70$ & $0.790 \pm 0.007$ & $0.763 \pm 0.020$ & $0.455 \pm 0.018$ \\
GradNorm & $44.35 \pm 1.12$ & $0.807 \pm 0.022$ & $0.802 \pm 0.003$ & $0.499 \pm 0.006$ \\
\textbf{RNN+OS} & $\mathbf{80.23 \pm 0.51}$ & $\mathbf{0.902 \pm 0.005}$ & $\mathbf{0.862 \pm 0.001}$ & $\mathbf{0.611 \pm 0.004}$ \\
\midrule
\multicolumn{5}{c}{K=5} \\
\midrule
RRR & $75.36 \pm 0.78$ & $0.828 \pm 0.009$ & $0.552 \pm 0.009$ & $0.068 \pm 0.019$ \\
LFADS & $87.49 \pm 0.50$ & $0.942 \pm 0.015$ & $0.907 \pm 0.009$ & $0.695 \pm 0.027$ \\
UW & $89.54 \pm 1.43$ & $0.920 \pm 0.014$ & $0.883 \pm 0.009$ & $0.633 \pm 0.032$ \\
GradNorm & $89.67 \pm 1.06$ & $0.919 \pm 0.003$ & $0.879 \pm 0.011$ & $0.622 \pm 0.029$ \\
\textbf{RNN+OS} & $\mathbf{89.68 \pm 0.41}$ & $\mathbf{0.952 \pm 0.004}$ & $\mathbf{0.922 \pm 0.002}$ & $\mathbf{0.735 \pm 0.010}$ \\
\bottomrule
\end{tabular}%
}
\caption{
Performance comparison with multi-task and neural decoding baselines on cross-session generalization under 2-shot and 5-shot settings (mean±std). Direction is reported as accuracy (\%), regression tasks (position, velocity, acceleration) reported as $R^2$ values. Bold indicates best performance per K-shot group.
}
\label{tab:cross_session_comparison}
\end{table}

\subsection{Experimental Results}

We conduct generalization experiments using CO and RT datasets on three representative neural decoding backbone architectures (CNN, RNN, and MLP) \citep{Liu2022}, evaluating performance across three scenarios: cross-session, cross-subject, and cross-paradigm. We use direction classification accuracy to evaluate motor intention decoding performance and $R^2$ values to assess motor regression for position, velocity, and acceleration under few-shot settings with $k \in \{2, 5\}$. Bolded values are the best results within each K-shot group. ``/'' represents results at chance level due to insufficient feature extraction capability. ``-'' indicates the task is not applicable. +OS indicates enhanced with OrthoSchema.

Table~\ref{tab:across_session} presents cross-session generalization performance on CO and RT paradigms. Results demonstrate that our method (+OS) achieves significant improvements across all metrics. Under the more restrictive 2-shot setting, performance improvements are more pronounced, validating the method's effectiveness under extremely limited sample conditions. Table~\ref{tab:across_subject} shows cross-subject generalization results. Due to differences in neuronal recording, and motor control strategies across different individuals, cross-subject transfer is more challenging than cross-session. All backbone networks showed sustained improvements, with MLP results omitted under k=2 conditions due to chance-level performance. Similarly, improvement effects are more pronounced under k=2 conditions across all architectures. Table~\ref{tab:across_task} presents cross-paradigm transfer results from CO to RT paradigms. Cross-paradigm transfer is the most challenging scenario due to different motor patterns, dynamics, and neural representations across paradigms. Nevertheless, our method still achieves clear improvements on most metrics, with more significant improvements under k=2 conditions.

\begin{figure}[t]
\centering
\includegraphics[width=\columnwidth]{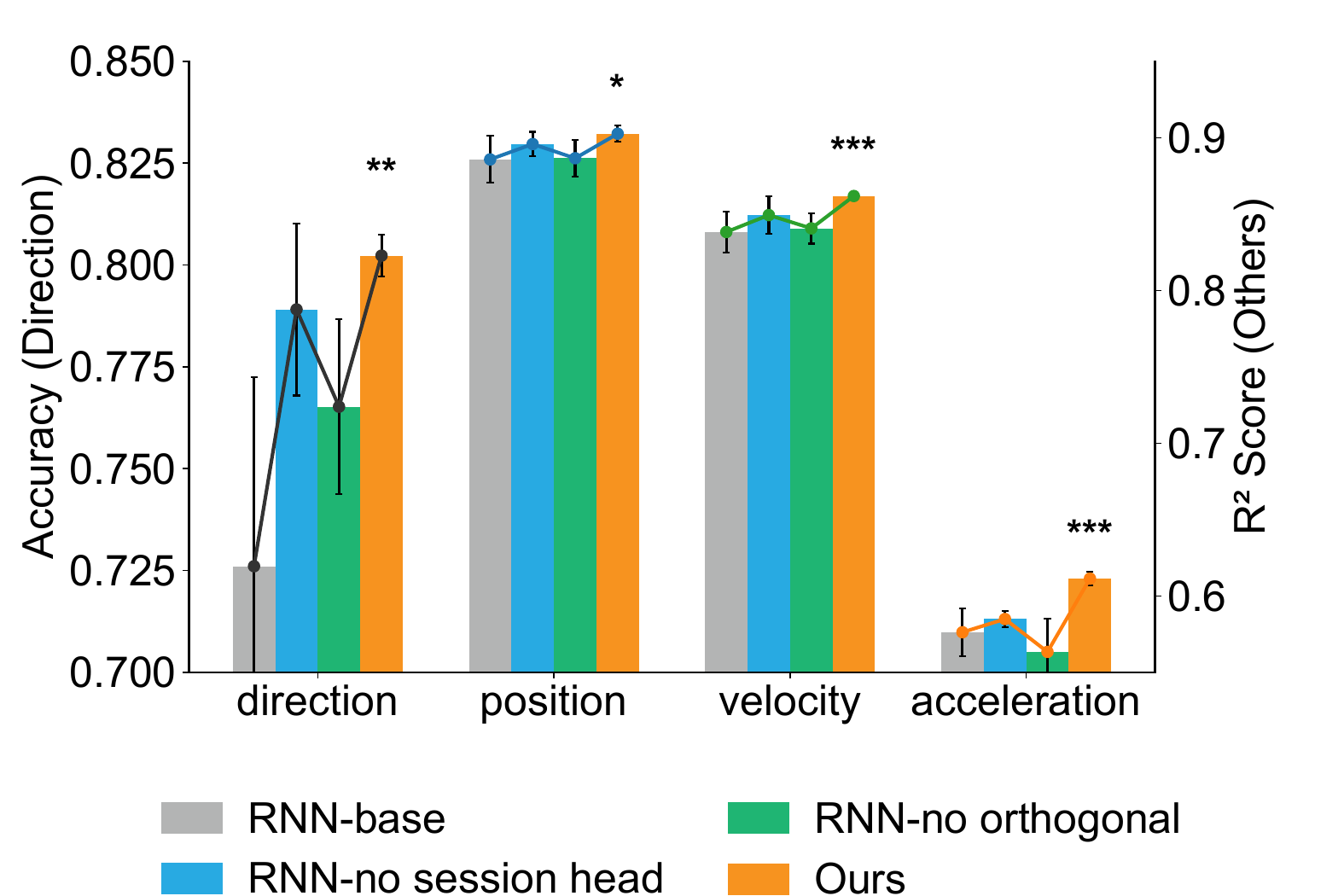} % Reduce the figure size so that it is slightly narrower than the column. Don't use precise values for figure width.This setup will avoid overfull boxes.
\caption{Performance comparison of different module combinations on cross-session generalization using RNN backbone under 2-shot setting.*, **, *** indicate statistical significance at $p<0.05$, $p<0.01$, $p<0.001$, respectively.}
\label{fig4}
\end{figure} 

\begin{figure}[t]
\centering
\includegraphics[width=\columnwidth]{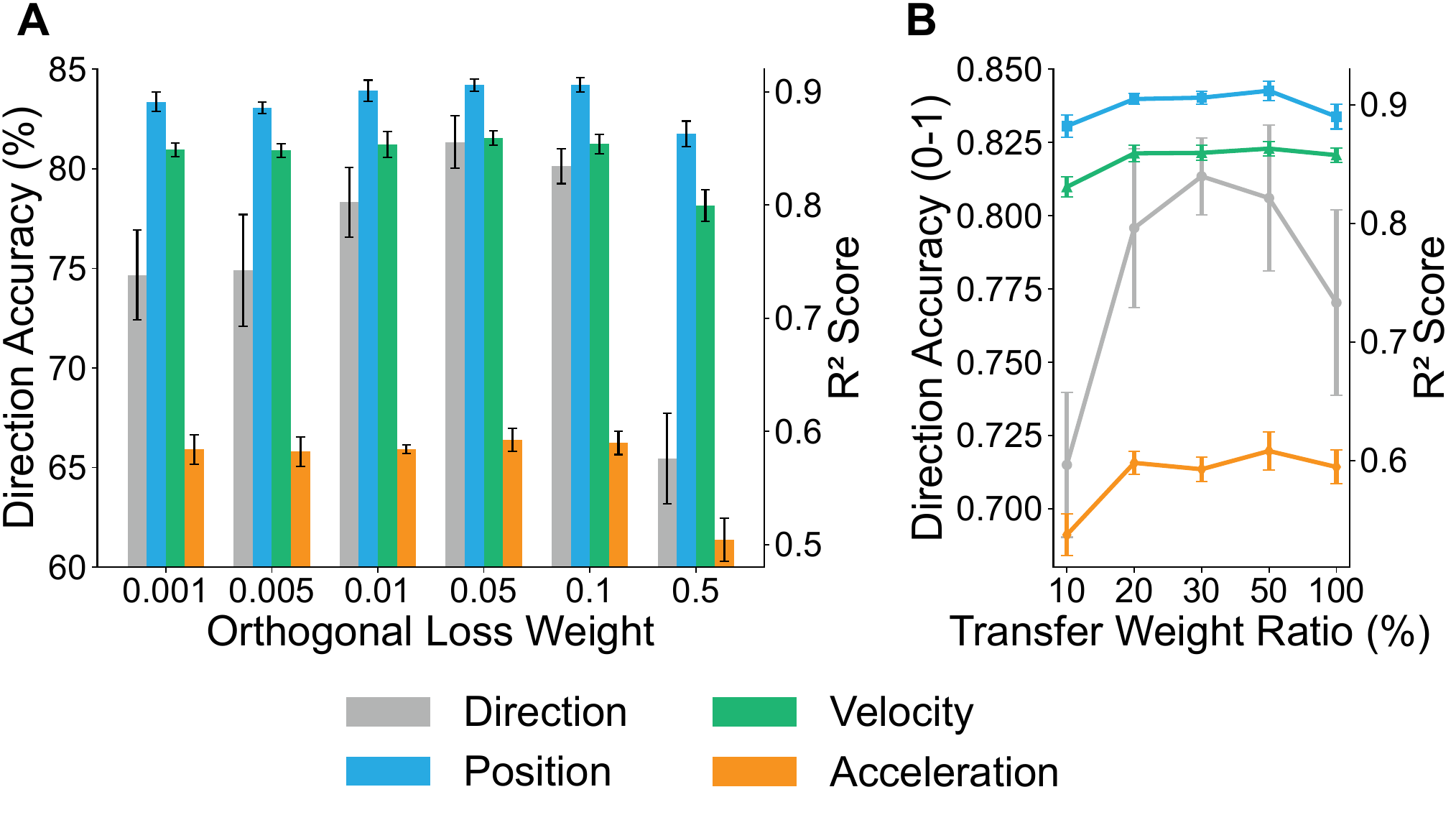} % Reduce the figure size so that it is slightly narrower than the column. Don't use precise values for figure width.This setup will avoid overfull boxes.
\caption{Hyperparameter analysis on cross-session generalization under 2-shot setting. (A) Impact of different orthogonal loss weights. (B) Impact of different feature retention ratios for transfer.}
\label{fig5}
\end{figure} 

We also compare OrthoSchema with classic multi-task and neural decoding baseline methods to validate its effectiveness in MND tasks. Table~\ref{tab:cross_session_comparison} shows cross-session performance comparisons on CO task with neuroscience decoding method Reduced Rank Regression (RRR) \citep{borgognon2025regional}, LFADS \citep{Pandarinath2018}, classic multi-task learning uncertainty weighting (UW) \citep{Kendall_2018_CVPR}, and gradient normalization (GradNorm) \citep{Chen2018GradNorm} methods. Our results outperform all methods, especially under k=2 conditions. OrthoSchema also maintains similar results for other cross-subject and cross-paradigm scenarios, with details provided in the appendix.

Overall, our method achieves significant improvements under both cross-session and cross-subject settings, and also demonstrates good enhancement effects in challenging cross-paradigm condition. The k-shot fine-tuning mechanism provides more significant performance improvements under extreme sample constraints (k=2), further demonstrating the effectiveness of our method.

\begin{figure}[t]
\centering
\includegraphics[width=\columnwidth]{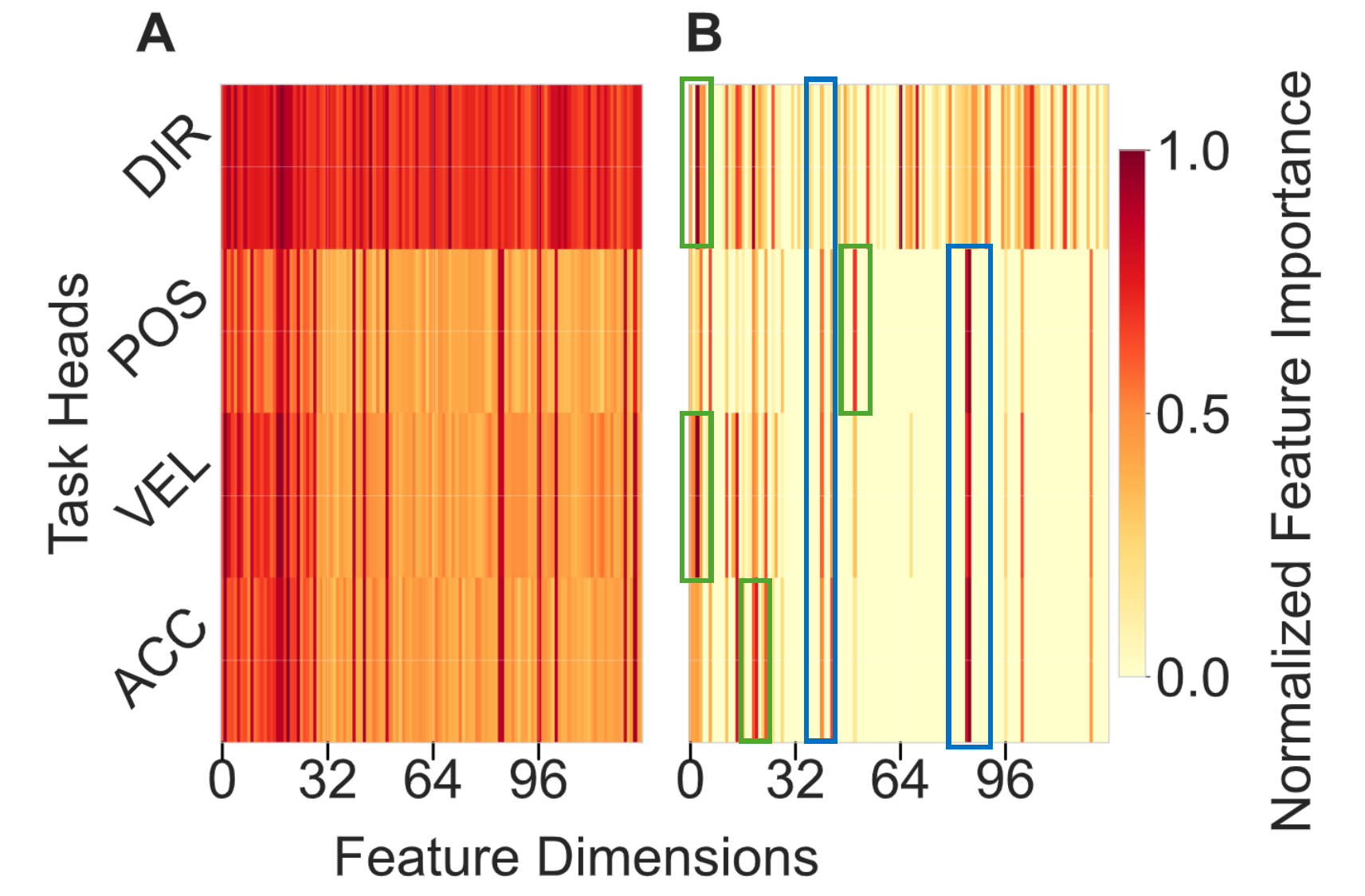} % Reduce the figure size so that it is slightly narrower than the column. Don't use precise values for figure width.This setup will avoid overfull boxes.
\caption{Comparison heatmap of feature importance extracted by orthogonality loss versus transferred feature importance. (A) Feature importance distribution of each head extracted by the trained network. (B) Adjusted feature importance transferred to new data after removing the influence of session/subject heads. Blue boxes indicate cross-task shared features, while green boxes represent task-specific features.}
\label{fig6}
\end{figure} 

\subsection{Ablation Studies}

\paragraph{Impact of Module Combinations} We first evaluate the impact of different module combinations on performance. The variant models are defined as follows: (i) Base multi-task RNN, containing only direction and motor regression tasks; (ii) Multi-task RNN with orthogonal loss; (iii) Multi-task RNN with session classification head; (iv) Ours, combining all components. As shown in Figure~\ref{fig4}, all components positively impact performance. Compared to the base model, adding orthogonal loss improves direction accuracy by approximately 6\%, adding session classification head brings about 2\% improvement, while the complete method achieves a significant 8\% enhancement. Statistical significance is also observed in regression tasks compared to other methods. This validates the effectiveness of each component and the importance of their synergistic effects.

\paragraph{Impact of Hyperparameter Settings} We analyze the impact of key hyperparameters to determine optimal configurations and understand model behavior. Figure~\ref{fig5}(A) shows the impact of orthogonal loss weights. Small weights (0.001-0.005) provide insufficient orthogonal constraints to decouple task features, while large weights (0.1-0.5) impose overly strong constraints that harm shared feature learning. A weight of 0.05 achieves optimal balance, ensuring feature decoupling while maintaining task performance. Figure~\ref{fig5}(B) analyzes feature retention ratios. Low ratios (10-20\%) cause loss of critical task information and performance degradation, while high ratios (50-100\%) retain excessive session-specific noise that affects generalization. A 30\% retention ratio achieves optimal performance by balancing information preservation and noise filtering.

\paragraph{Extended Validation Experiments} To further validate the robustness of our method, we conducted two extended experiments. First, the impact of training days (Appendix Figure 3) shows that more training days lead to better performance, especially pronounced in direction tasks, demonstrating that the model can effectively utilize more training data to learn robust neural-behavioral mapping relationships. Second, single-task decoding validation (Appendix Figure 4) shows that our model framework is also effective in single-task decoding, confirming the method's broad applicability. Detailed results are provided in the appendix.

\begin{figure}[t]
\centering
\includegraphics[width=0.8\columnwidth]{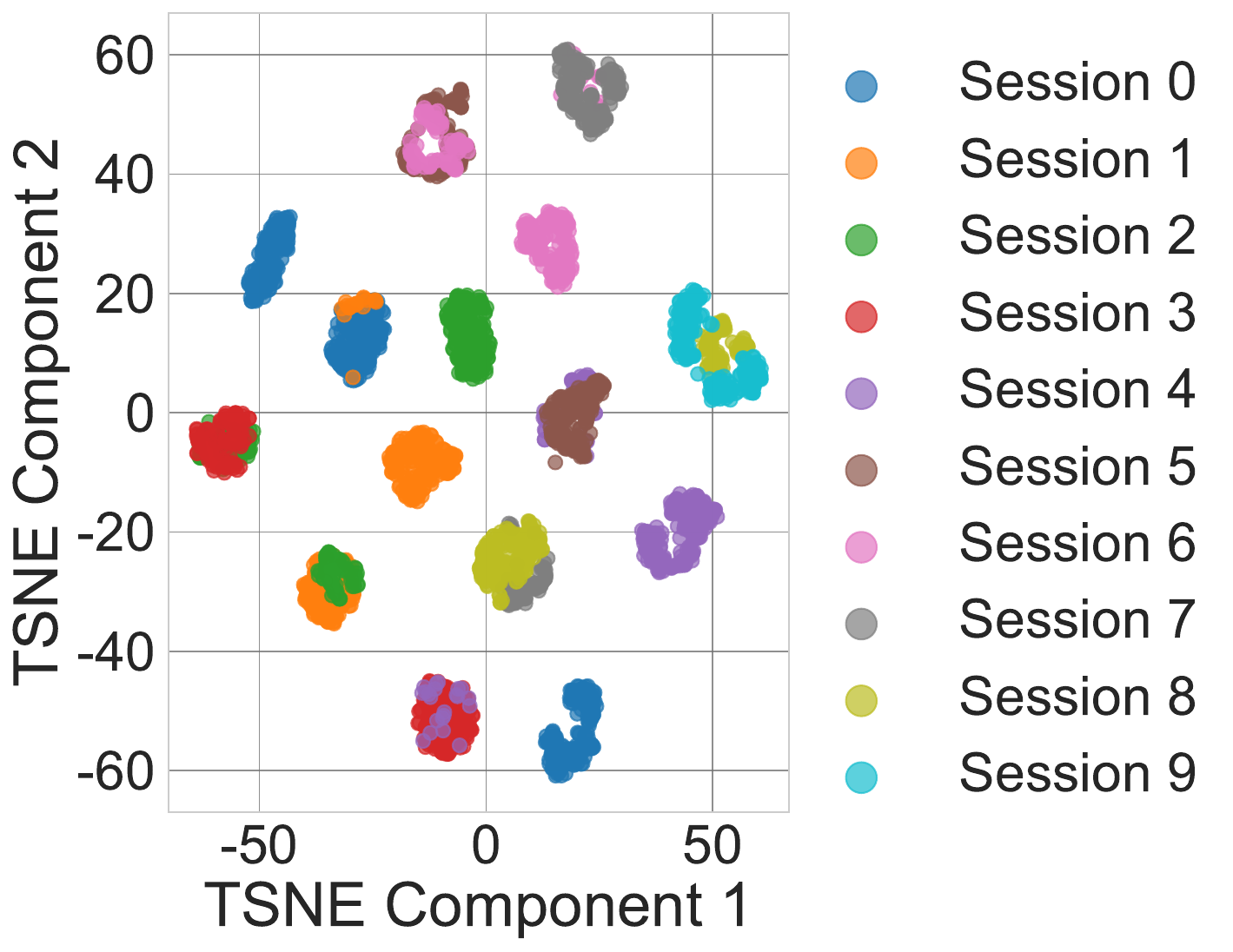} % Reduce the figure size so that it is slightly narrower than the column. Don't use precise values for figure width.This setup will avoid overfull boxes.
\caption{T-SNE projection of the learned session embeddings. Different colors represent different sessions.}
\label{fig7}
\end{figure} 

\subsection{Visualization Analysis}
To provide insights into our method's internal mechanisms, we conduct feature importance visualization analysis. Figure~\ref{fig6} demonstrates the model's effectiveness in feature decomposition and transfer. Figure ~\ref{fig6} (A) shows that different task heads learn distinct feature importance patterns, proving successful cross-task feature disentanglement. Figure ~\ref{fig6} (B) displays the adjusted feature importance after removing day/subject head influences. The transferred patterns are more concentrated and stable, with shared feature regions (blue boxes) showing stronger consistency while preserving task-specific features (green boxes). The transferred feature distribution exhibits less noise and clearer structure, directly explaining the method's significant improvements in cross-scenario generalization. We further visualize the learned representations to observe how the model learns inter-session relationships (Figure~\ref{fig7}). Analysis shows that adjacent sessions map to similar regions in low-dimensional space, indicating that the model not only learns to identify task-specific and shared patterns, but also captures macro-structural relationships across multiple sessions.

\section{Conclusion}
This work defines the MND task and proposes the \textbf{OrthoSchema} framework as the solution, eliminating cross-task interference through orthogonal constraints and achieving cross-session, subjects and paradigms few-shot adaptation through schema-like selective feature transfer. Experiments on macaque motor cortex data validate the effectiveness of OrthoSchema, particularly under limited data conditions (K=2), with visualization revealing effective orthogonal disentanglement and schema-based transfer. This work provides insights into multi-dimensional neural information extraction and generalization, informing the development of high-bandwidth BCIs.

\bibliography{aaai2026}

\begin{thebibliography}{43}
\providecommand{\natexlab}[1]{#1}

\bibitem[{Alamgir, Grosse-Wentrup, and Altun(2010)}]{Alamgir2010}
Alamgir, M.; Grosse-Wentrup, M.; and Altun, Y. 2010.
\newblock Multitask learning for brain-computer interfaces.
\newblock In \emph{Proceedings of the Thirteenth International Conference on Artificial Intelligence and Statistics}, 17--24.

\bibitem[{Bernardi et~al.(2020)Bernardi, Benna, Rigotti, Munuera, Fusi, and Salzman}]{Bernardi2020}
Bernardi, S.; Benna, M.~K.; Rigotti, M.; Munuera, J.; Fusi, S.; and Salzman, C.~D. 2020.
\newblock The Geometry of Abstraction in the Hippocampus and Prefrontal Cortex.
\newblock \emph{Cell}, 183(4): 954--967.e21.
\newblock Open Archive.

\bibitem[{Bertalan and Hess(2021)}]{Bertalan2021}
Bertalan, B.; and Hess, G.~A. 2021.
\newblock Multi-Task Learning for Deep Neural Network Representations of Human Brain Activity.
\newblock \emph{arXiv preprint}.

\bibitem[{Borgognon et~al.(2025)Borgognon, Macellari, Hickey et~al.}]{borgognon2025regional}
Borgognon, S.; Macellari, N.; Hickey, A.~M.; et~al. 2025.
\newblock Regional specialization of movement encoding across the primate sensorimotor cortex.
\newblock \emph{Nature Communications}, 16: 5729.

\bibitem[{Chen et~al.(2018)Chen, Badrinarayanan, Lee, and Rabinovich}]{Chen2018GradNorm}
Chen, Z.; Badrinarayanan, V.; Lee, C.-Y.; and Rabinovich, A. 2018.
\newblock GradNorm: Gradient Normalization for Adaptive Loss Balancing in Deep Multitask Networks.
\newblock In \emph{Proceedings of the 35th International Conference on Machine Learning (ICML)}, volume~80, 794--803. PMLR.

\bibitem[{Chestek et~al.(2011)Chestek, Batista, Santhanam, Yu, Afshar, Cunningham, and Shenoy}]{Chestek2011}
Chestek, C.~A.; Batista, A.~P.; Santhanam, G.; Yu, B.~M.; Afshar, A.; Cunningham, J.~P.; and Shenoy, K.~V. 2011.
\newblock Single-neuron stability during repeated reaching in macaque premotor cortex.
\newblock \emph{Journal of Neuroscience}, 31(46): 17235--17243.

\bibitem[{Collinger et~al.(2013)Collinger, Wodlinger, Downey, Wang, Tyler-Kabara, Weber, and Schwartz}]{Collinger2013}
Collinger, J.~L.; Wodlinger, B.; Downey, J.~E.; Wang, W.; Tyler-Kabara, E.~C.; Weber, D.~J.; and Schwartz, A.~B. 2013.
\newblock High-performance neuroprosthetic control by an individual with tetraplegia.
\newblock \emph{The Lancet}, 381(9866): 557--564.

\bibitem[{Cunningham and Yu(2014)}]{Cunningham2014}
Cunningham, J.~P.; and Yu, B.~M. 2014.
\newblock Dimensionality reduction for large-scale neural recordings.
\newblock \emph{Nature Neuroscience}, 17(11): 1500--1509.

\bibitem[{Degenhart et~al.(2020)Degenhart, Bishop, Oby, Tyler-Kabara, Chase, Batista, and Byron}]{Degenhart2020}
Degenhart, A.~D.; Bishop, W.~E.; Oby, E.~R.; Tyler-Kabara, E.~C.; Chase, S.~M.; Batista, A.~P.; and Byron, M.~Y. 2020.
\newblock Stabilization of a brain--computer interface via the alignment of low-dimensional spaces of neural activity.
\newblock \emph{Nature Biomedical Engineering}, 4(7): 672--685.

\bibitem[{Dong et~al.(2023)Dong, Wu, Xiong, Li, Cheng, He, Qian, Cao, and Mo}]{dong2023gdod}
Dong, X.; Wu, R.; Xiong, C.; Li, H.; Cheng, L.; He, Y.; Qian, S.; Cao, J.; and Mo, L. 2023.
\newblock GDOD: Effective Gradient Descent using Orthogonal Decomposition for Multi-Task Learning.
\newblock \emph{arXiv preprint arXiv:2301.13465}.

\bibitem[{Elsayed et~al.(2016)Elsayed, Lara, Kaufman, Churchland, and Cunningham}]{Elsayed2016}
Elsayed, G.~F.; Lara, A.~H.; Kaufman, M.~T.; Churchland, M.~M.; and Cunningham, J.~P. 2016.
\newblock Reorganization between preparatory and movement population responses in motor cortex.
\newblock \emph{Nature Communications}, 7(1): 13239.

\bibitem[{Farshchian et~al.(2018)Farshchian, Gallego, Cohen, Bengio, Miller, and Solla}]{farshchian2018adversarial}
Farshchian, A.; Gallego, J.~A.; Cohen, J.~P.; Bengio, Y.; Miller, L.~E.; and Solla, S.~A. 2018.
\newblock Adversarial domain adaptation for stable brain-machine interfaces.
\newblock \emph{arXiv preprint arXiv:1810.00045}.

\bibitem[{Flesch et~al.(2022)Flesch, Juechems, Dumbalska, Saxe, and Summerfield}]{Flesch2022}
Flesch, T.; Juechems, K.; Dumbalska, T.; Saxe, A.; and Summerfield, C. 2022.
\newblock Orthogonal representations for robust context-dependent task performance in brains and neural networks.
\newblock \emph{Neuron}, 110(7): 1258--1270.e11.
\newblock Open access.

\bibitem[{Flesher et~al.(2021)Flesher, Downey, Weiss, Hughes, Herrera, Tyler-Kabara, and Collinger}]{Flesher2021}
Flesher, S.~N.; Downey, J.~E.; Weiss, J.~M.; Hughes, C.~L.; Herrera, A.~J.; Tyler-Kabara, E.~C.; and Collinger, J.~L. 2021.
\newblock A brain-computer interface that evokes tactile sensations improves robotic arm control.
\newblock \emph{Science}, 372(6544): 831--836.

\bibitem[{Gallego, Makin, and McDougle(2022)}]{Gallego2022}
Gallego, J.~A.; Makin, T.~R.; and McDougle, S.~D. 2022.
\newblock Going beyond primary motor cortex to improve brain--computer interfaces.
\newblock \emph{Trends in Neurosciences}, 45(3): 176--183.

\bibitem[{Georgopoulos, Schwartz, and Kettner(1986)}]{Georgopoulos1986}
Georgopoulos, A.~P.; Schwartz, A.~B.; and Kettner, R.~E. 1986.
\newblock Neuronal population coding of movement direction.
\newblock \emph{Science}, 233(4771): 1416--1419.

\bibitem[{Glaser et~al.(2020)Glaser, Benjamin, Chowdhury, Perich, Miller, and Kording}]{Glaser2020}
Glaser, J.~I.; Benjamin, A.~S.; Chowdhury, R.~H.; Perich, M.~G.; Miller, L.~E.; and Kording, K.~P. 2020.
\newblock Machine learning for neural decoding.
\newblock \emph{eNeuro}, 7(4).

\bibitem[{Goudar et~al.(2023)Goudar, Peysakhovich, Freedman, Buffalo, and Wang}]{Goudar2023Schema}
Goudar, V.; Peysakhovich, B.; Freedman, D.~J.; Buffalo, E.~A.; and Wang, X.-J. 2023.
\newblock Schema formation in a neural population subspace underlies learning-to-learn in flexible sensorimotor problem-solving.
\newblock \emph{Nature Neuroscience}, 26(5): 879--890.

\bibitem[{Hochberg et~al.(2012)Hochberg, Bacher, Jarosiewicz, Masse, Simeral, Vogel, and Donoghue}]{Hochberg2012}
Hochberg, L.~R.; Bacher, D.; Jarosiewicz, B.; Masse, N.~Y.; Simeral, J.~D.; Vogel, J.; and Donoghue, J.~P. 2012.
\newblock Reach and grasp by people with tetraplegia using a neurally controlled robotic arm.
\newblock \emph{Nature}, 485(7398): 372--375.

\bibitem[{Karpowicz et~al.(2025)Karpowicz, Ali, Wimalasena, Sedler, Shanechi, Batista, and Pandarinath}]{Karpowicz2025}
Karpowicz, B.~M.; Ali, Y.~H.; Wimalasena, L.~N.; Sedler, A.~R.; Shanechi, M.~M.; Batista, A.~P.; and Pandarinath, C. 2025.
\newblock Stabilizing brain-computer interfaces through alignment of latent dynamics.
\newblock \emph{Nature Communications}, 16(1): 1--17.

\bibitem[{Kaufman et~al.(2014)Kaufman, Churchland, Ryu, and Shenoy}]{Kaufman2014}
Kaufman, M.~T.; Churchland, M.~M.; Ryu, S.~I.; and Shenoy, K.~V. 2014.
\newblock Cortical activity in the null space: permitting preparation without movement.
\newblock \emph{Nature Neuroscience}, 17(3): 440--448.

\bibitem[{Kendall, Gal, and Cipolla(2018)}]{Kendall_2018_CVPR}
Kendall, A.; Gal, Y.; and Cipolla, R. 2018.
\newblock Multi-Task Learning Using Uncertainty to Weigh Losses for Scene Geometry and Semantics.
\newblock In \emph{Proceedings of the IEEE Conference on Computer Vision and Pattern Recognition (CVPR)}.

\bibitem[{Li et~al.(2024)Li, Zhu, Qi, and Wang}]{10.7554/eLife.87881}
Li, Y.; Zhu, X.; Qi, Y.; and Wang, Y. 2024.
\newblock Revealing unexpected complex encoding but simple decoding mechanisms in motor cortex via separating behaviorally relevant neural signals.
\newblock \emph{eLife}, 12: RP87881.

\bibitem[{Li(2014)}]{Li2014}
Li, Z. 2014.
\newblock Decoding methods for neural prostheses: where have we reached?
\newblock \emph{Frontiers in Systems Neuroscience}, 8: 129.

\bibitem[{Lim et~al.(2025)Lim, Lo, Tan, Lin, Wang, Tan, Wang, Ma, Ng, and Jeffree}]{Lim2025}
Lim, M. J.~R.; Lo, J. Y.~T.; Tan, Y.~Y.; Lin, H.-Y.; Wang, Y.; Tan, D.; Wang, E.; Ma, Y. Y.~N.; Ng, J. J.~W.; and Jeffree, R.~A. 2025.
\newblock The state-of-the-art of invasive brain-computer interfaces in humans: a systematic review and individual patient meta-analysis.
\newblock \emph{Journal of Neural Engineering}, 22(2): 026013.

\bibitem[{Liu et~al.(2022)Liu, Meamardoost, Gunawan, Komiyama, Mewes, Zhang, and Wang}]{Liu2022}
Liu, F.; Meamardoost, S.; Gunawan, R.; Komiyama, T.; Mewes, C.; Zhang, Y.; and Wang, L. 2022.
\newblock Deep learning for neural decoding in motor cortex.
\newblock \emph{Journal of Neural Engineering}, 19(5): 056021.

\bibitem[{Musallam et~al.(2004)Musallam, Corneil, Greger, Scherberger, and Andersen}]{Musallam2004}
Musallam, S.; Corneil, B.~D.; Greger, B.; Scherberger, H.; and Andersen, R.~A. 2004.
\newblock Cognitive control signals for neural prosthetics.
\newblock \emph{Science}, 305(5681): 258--262.

\bibitem[{Oikonomou, Nikolopoulos, and Kompatsiaris(2022)}]{Oikonomou2022}
Oikonomou, V.~P.; Nikolopoulos, S.; and Kompatsiaris, I. 2022.
\newblock A multitask bayesian framework for the analysis of motor imagery eeg data.
\newblock In \emph{2022 30th European Signal Processing Conference (EUSIPCO)}, 1308--1312. IEEE.

\bibitem[{Pandarinath et~al.(2018)Pandarinath, O'Shea, Collins, Jozefowicz, Stavisky, Kao, and Sussillo}]{Pandarinath2018}
Pandarinath, C.; O'Shea, D.~J.; Collins, J.; Jozefowicz, R.; Stavisky, S.~D.; Kao, J.~C.; and Sussillo, D. 2018.
\newblock Inferring single-trial neural population dynamics using sequential auto-encoders.
\newblock \emph{Nature Methods}, 15(10): 805--815.

\bibitem[{Perge et~al.(2013)Perge, Homer, Malik, Cash, Eskandar, Friehs, and Hochberg}]{Perge2013}
Perge, J.~A.; Homer, M.~L.; Malik, W.~Q.; Cash, S.; Eskandar, E.; Friehs, G.; and Hochberg, L.~R. 2013.
\newblock Intra-day signal instabilities affect decoding performance in an intracortical neural interface system.
\newblock \emph{Journal of Neural Engineering}, 10(3): 036004.

\bibitem[{Perich et~al.(2025)Perich, Miller, Azabou, and Dyer}]{Perich2025}
Perich, M.~G.; Miller, L.~E.; Azabou, M.; and Dyer, E.~L. 2025.
\newblock Long-term recordings of motor and premotor cortical spiking activity during reaching in monkeys (Version 0.250122.1735) [Data set].
\newblock DANDI Archive.
\newblock \url{https://doi.org/10.48324/dandi.000688/0.250122.1735}.

\bibitem[{Samborska et~al.(2022)Samborska, Butler, Walton, Behrens, and Akam}]{samborska2022complementary}
Samborska, V.; Butler, J.~L.; Walton, M.~E.; Behrens, T. E.~J.; and Akam, T. 2022.
\newblock Complementary task representations in hippocampus and prefrontal cortex for generalizing the structure of problems.
\newblock \emph{Nature Neuroscience}, 25(10): 1314--1326.

\bibitem[{Shih, Krusienski, and Wolpaw(2012)}]{Shih2012}
Shih, J.~J.; Krusienski, D.~J.; and Wolpaw, J.~R. 2012.
\newblock Brain-computer interfaces in medicine.
\newblock In \emph{Mayo Clinic Proceedings}, volume~87, 268--279. Elsevier.

\bibitem[{Tang et~al.(2020)Tang, Herikstad, Parthasarathy, Libedinsky, and Yen}]{Tang2020Minimally}
Tang, C.; Herikstad, R.; Parthasarathy, A.; Libedinsky, C.; and Yen, S.-C. 2020.
\newblock Minimally dependent activity subspaces for working memory and motor preparation in the lateral prefrontal cortex.
\newblock \emph{eLife}, 9: e58154.

\bibitem[{Tian et~al.(2024)Tian, Zhao, Chen, Wang, Qi, Constantinidis, and Zhou}]{Tian2024}
Tian, K.; Zhao, Z.; Chen, Y.; Wang, X.; Qi, X.~L.; Constantinidis, C.; and Zhou, X. 2024.
\newblock Domain-specific Schema Reuse Supports Flexible Learning to Learn in Primate Brain.
\newblock \emph{bioRxiv}.
\newblock Preprint.

\bibitem[{Willett et~al.(2021)Willett, Avansino, Hochberg, Henderson, and Shenoy}]{Willett2021}
Willett, F.~R.; Avansino, D.~T.; Hochberg, L.~R.; Henderson, J.~M.; and Shenoy, K.~V. 2021.
\newblock High-performance brain-to-text communication via handwriting.
\newblock \emph{Nature}, 593(7858): 249--254.

\bibitem[{Wimalasena, Miller, and Pandarinath(2020)}]{Wimalasena2020}
Wimalasena, L.~N.; Miller, L.~E.; and Pandarinath, C. 2020.
\newblock From unstable input to robust output.
\newblock \emph{Nature Biomedical Engineering}, 4(7): 665--667.

\bibitem[{Wolpaw(2013)}]{Wolpaw2013}
Wolpaw, J.~R. 2013.
\newblock Brain--computer interfaces.
\newblock In \emph{Handbook of Clinical Neurology}, volume 110, 67--74. Elsevier.

\bibitem[{Wu et~al.(2006)Wu, Gao, Bienenstock, Donoghue, and Black}]{Wu2006}
Wu, W.; Gao, Y.; Bienenstock, E.; Donoghue, J.~P.; and Black, M.~J. 2006.
\newblock Bayesian population decoding of motor cortical activity using a Kalman filter.
\newblock \emph{Neural Computation}, 18(1): 80--118.

\bibitem[{Xie, Schwartz, and Prasad(2018)}]{Xie2018}
Xie, Z.; Schwartz, O.; and Prasad, A. 2018.
\newblock Decoding of finger trajectory from ECoG using deep learning.
\newblock \emph{Journal of Neural Engineering}, 15(3): 036009.

\bibitem[{Ye et~al.(2024)Ye, Collinger, Gaunt, Dyer, and Batista}]{Ye2024a}
Ye, J.; Collinger, J.~L.; Gaunt, R.~A.; Dyer, E.~L.; and Batista, A.~P. 2024.
\newblock POYO: A neural data transformer for multi-session population dynamics.
\newblock \emph{Nature Machine Intelligence}, 6(3): 287--301.

\bibitem[{Ye et~al.(2023)Ye, Pandarinath, Versteeg, Sedler, Chowdhury, Kaufman, and Dyer}]{Ye2023}
Ye, J.; Pandarinath, C.; Versteeg, C.; Sedler, A.~R.; Chowdhury, R.~H.; Kaufman, M.~T.; and Dyer, E.~L. 2023.
\newblock Neural data transformer for large-scale multi-session neural population activity inference.
\newblock arXiv preprint.
\newblock ArXiv:2305.10812.

\bibitem[{Zhou et~al.(2021)Zhou, Jia, Montesinos‑Cartagena, Zhang, Musall, Nagel, and Namboodiri}]{Zhou2021}
Zhou, J.; Jia, C.; Montesinos‑Cartagena, M.; Zhang, S.; Musall, S.; Nagel, K.~I.; and Namboodiri, V. M.~K. 2021.
\newblock Evolving schema representations in orbitofrontal ensembles during learning.
\newblock \emph{Nature}, 590(7847): 606--611.

\end{thebibliography}

\end{document}